\documentclass[a4paper, 11pt]{article}
\usepackage[left=1.5cm, right=1.5cm, bottom = 2cm, top=2.8cm]{geometry}
\usepackage{amsmath, amssymb}
\usepackage{graphicx}
\usepackage[russian, english]{babel}
\usepackage{enumitem}
\usepackage{titling}
\usepackage{xfrac}
\usepackage{hyperref}
\usepackage{float}
\usepackage{bm}
\hypersetup{				
    unicode=true,           
    pdftitle={DLU 2019 Presentation},
    pdfauthor={R. I. Gainutdinov, Yu. V. Baryshev, V. V. Sokolov},
    pdfsubject={S-stars motion around RCO Sgr A*},
    pdfcreator={},
    pdfproducer={SPbSU, SAO RAS},
    pdfkeywords={GR} {FGT} {Gravitation theory} {Sgr A*} {Post-newtonian approximation},
    colorlinks=false,       	
    linkcolor=blue,          
    citecolor=green,        
    filecolor=magenta,      
    urlcolor=cyan           
}

\usepackage{listings}
\usepackage{makecell}
\usepackage{atbegshi,picture}
\usepackage{xcolor}

\title{\textbf{S-stars motion around relativistic compact object Sgr A*.}}
\author{R. I. Gainutdinov$^{\text{1}}$, Yu. V. Baryshev$^{\text{1}}$, V. V. Sokolov$^{\text{2}}$.\\\\
$^{\text{1}}$\emph{Saint Petersburg State University}, $^{\text{2}}$\emph{SAO RAS}.}
\date{DLU 2019 SAO RAS, 03.10.2019}
\begin{document}

\maketitle

\sloppy

\section*{Abstract}
    A review of modern VLTI observations of the orbital motion of closest stars to the relativistic compact object Sgr A* and its ability to test gravitation theories in the conditions of Post-Newtonian approximation. The observed orbital parameters, second order Doppler effect and gravitational redshift, measured for several S-stars, are compared with theoretical PN predictions.

\tableofcontents

\section{Introduction}
    \subsection{Review of modern observations of S-stars}
        For many years there exists a demand for any observational data that can give a good opportunity to investigate different effects and phenomenas predicted by different gravitation theories to compare them. Modern observations of the cluster of stars rotating around the relativistic compact object Sgr A* in the center of our Galaxy, known as ``S-stars", provide us with such an opportunity. Having tight orbits with close distance to the black hole and very high velocities during periapse passage, they can give us a good experimental data for obtaining orbital parameters, pericenter precession, Doppler effect and gravitational redshift. This data was used to find the distance and mass of Sgr A* (\cite{Keck}, \cite{G}). They are:\\
        \[M=4.28\cdot10^6 M_\odot\]
        \[R_g=\frac{GM}{c^2}=6.32\cdot 10^{11} \textrm{ cm}\]
        \[D = 8.32 \textrm{ kpc}\]
        And these parameters also could be used to check the predictions of several gravitation theories in terms of Post-Newtonian approximation.\\
The stars that we are interested in the most are three closest to Sgr A* stars: S2 (also known as S0-2), S38 and S55 (also known as S0-102).
        \begin{table}[H]
            \centering
            \begin{tabular}{lcccc}
                \hline
                Star & $a\textrm{, A.U.}$ & $e$ & $T\textrm{, yrs}$ & $v_p\textrm{, km/s}$ \\ \hline
                S2 (S0-2)    & $1044$ & $0.8839$ & $16.0$ & $7675$ \\
                S38 (S0-38)  & $1178$ & $0.8201$ & $19.2$ & $5709$ \\
                S55 (S0-102) & $897$  & $0.7209$ & $12.8$ & $5108$ \\ \hline
            \end{tabular}
            \caption{S-stars orbital parameters}\label{tab:Sstars}
        \end{table}
        \noindent The star S2 passed its pericenter in 2002 and in 2018, so its orbital elements could be obtained from observations with relatively high accuracy. During the May 2018 passage it reached the velocity of $\sim$ 0.02 the speed of light, providing the basis for gravitational redshift research (Tuan Do, \cite{S2redshift}, 2019). The star S55 is known as the star that has the shortest orbital period known, $\sim$ 12 years.\\
        The S-star cluster was investigated by Eisenhauer et al. \cite{S-astroph} in 2005. They have obtained some of the astrophysical parameters of S-stars:
        \begin{itemize}
            \item The magnitude of observed S-stars in $K$-band is $\sim 14^m\div16^m$.
            \item The observed S-stars have \emph{normal rotating velocities}, similar to solar neighborhood stars.
            \item The majority of S-stars appear to be a main sequence stars of $\textrm{B}0\div \textrm{B}9$ spectral classes.
        \end{itemize}
        The G2 cloud is also an interesting object. Because G2 was firstly considered as a cloud, it was predicted, that this object would not survive its pericenter passage due to tidal forces. But there was no tidal disruption, thus now there are reasons to assume that it is rather a star than a cloud. Observational data of G2 was used to find out its orbit parameters and gravitational redshift effect (S. Gillessen, P.M. Plewa, \cite{G2}, 2018).\\
        In the work of Leor Barack et al. (which has more than 200 authors) ``Black Holes, Gravitational Waves and Fundamental Physics: a roadmap" \cite{Barack}, which is an overview of relevant fields of research in fundamental physics, it is noted, that the astrophysical tests of the gravity physics: \emph{"all involve gravity as a key component"}.\\
        The S2 star data was used to constrain scalar-tensor gravity models (Vesna Borka Jovanović et al., \cite{j2019}). \emph{"The aim of our investigation is to derive a particular theory among the class
of \textbf{scalar-tensor(ST) theories of gravity}, and then to test it by studying kinematics and dynamics of S-stars around supermassive black hole (BH) at Galactic Center (GC)"}

    \subsection{Post-Newtonian Parameters}
    We consider a possibility for Post-Newtonian  measurement of the \textbf{energy density of the gravitational field} using orbital motion of S-stars.\\
    The Post-Newtonian parameters are:
    \[\frac{v}{c}, \quad \frac{v^2}{c^2}, \quad \frac{\varphi_N}{c^2}, \quad z_d, \quad z_g, \quad \varepsilon_g\]
    \begin{itemize}
        \item $v$ -- orbital velocity;
        \item $\varphi_N = -GM/r$ -- Newtonian potential;
        \item $z_d$ -- relativistic Doppler shift;
        \item $z_g$ -- gravitational redshift;
        \item $\varepsilon_g$ -- energy density of the gravitational field (erg/cm$^\text{3}$).
    \end{itemize}$\;$\\
    We consider the contribution of the energy density of the gravitational field to the observed value of the pericenter shift of the S stars (S2, S38, S55).

\section{Post-Newonian effects}
\subsection{In General Relativity}
    In this section we will consider a problem of a massless test particle in a gravitational field of central massive body using Post-Newtonian approximation.
    \subsubsection{The equations of motion (GR)}
        In the frame of geometrical gravitaion theory (GR), according to Brumberg (\cite{Brumberg91}, 1991), the Lagrange function and corresponding Post-Newtonian equations of motion of a test particle in the static gravitational field of central massive body are given by:
        \begin{equation*}
            \frac{L}{m} = \frac{\dot{\mathbf{r}}^2}{2} \bigg( 1 + \frac{\dot{\mathbf{r}}^2}{4c^2} - (3-2\alpha)\frac{\varphi_N}{c^2} \bigg) - \varphi_N \bigg( 1 + (1 - 2\alpha)\frac{\varphi_N}{2c^2} + \alpha \frac{(\mathbf{r}\cdot\dot{\mathbf{r}})^2}{c^2r^2} \bigg)
        \end{equation*}
        \begin{equation*}
            \boxed{\ddot{\mathbf{r}} =-\bm{\nabla}\varphi_N \bigg( 1 + (4-2\alpha)\frac{\varphi_N}{c^2} + (1+\alpha)\frac{\dot{\mathbf{r}}^2}{c^2} - 3\alpha\frac{(\mathbf{r}\cdot\dot{\mathbf{r}})^2}{c^2r^2} \bigg) + (4-2\alpha)\bigg( \bm{\nabla}\varphi_N \cdot\frac{\dot{\mathbf{r}}}{c} \bigg)\frac{\dot{\mathbf{r}}}{c} }
        \end{equation*}
        From the Lagrange function we also obtain energy ($E = (\frac{\partial L}{\partial \dot{\mathbf{r}}}\cdot\dot{\mathbf{r}})-L$) and angular momentum ($\mathbf{J} = \mathbf{r} \times \frac{\partial L}{\partial \dot{\mathbf{r}}}$):
        \begin{equation*}
            \frac{E}{m} = \frac{\dot{\mathbf{r}}^2}{2} \bigg( 1 + \frac{\dot{\mathbf{r}}^2}{4c^2} - (3-2\alpha)\frac{\varphi_N}{c^2} \bigg) + \varphi_N \bigg( 1 + (1 - 2\alpha)\frac{\varphi_N}{2c^2} - \alpha \frac{(\mathbf{r}\cdot\dot{\mathbf{r}})^2}{c^2r^2} \bigg)
        \end{equation*}
        \begin{equation*}
            \frac{\mathbf{J}}{m} = [\mathbf{r}\times\dot{\mathbf{r}}]\bigg( 1 + \frac{\dot{\mathbf{r}}^2}{4c^2} - (3-2\alpha)\frac{\varphi_N}{c^2} \bigg)
        \end{equation*}
        Where $(\alpha=1)$ for Schwarzschild coordinates, and $(\alpha=0)$ for \textbf{harmonic} coordinates.\\
        The equations of motion in harmonic coordinates:
        \begin{equation}
        \label{GR-harmon}
            \boxed{ \ddot{\mathbf{r}}=-\bm{\nabla}\varphi_N \bigg( 1 + 4\frac{\varphi_N}{c^2} + \frac{\dot{\mathbf{r}}^2}{c^2}\bigg) + 4\bigg( \bm{\nabla}\varphi_N \cdot\frac{\dot{\mathbf{r}}}{c} \bigg)\frac{\dot{\mathbf{r}}}{c} }
        \end{equation}
        The energy and the angular momentum in harmonic coordinates:
        \begin{equation*}
            \frac{E}{m} = \frac{\dot{\mathbf{r}}^2}{2} \bigg( 1 + \frac{\dot{\mathbf{r}}^2}{4c^2} - 3\frac{\varphi_N}{c^2} \bigg) + \varphi_N \bigg( 1 + \frac{\varphi_N}{2c^2} \bigg)
        \end{equation*}
        \begin{equation*}
            \frac{\mathbf{J}}{m} = [\mathbf{r}\times\dot{\mathbf{r}}]\bigg( 1 + \frac{\dot{\mathbf{r}}^2}{4c^2} - 3\frac{\varphi_N}{c^2} \bigg)
        \end{equation*}
        
    \subsubsection{Pericenter shift (GR)}
        The Post-Newtonian equation of motion lead us into the effect of pericenter shift. The formula for the value of pericenter shift per one turn in GR is well known (\cite{Brumberg91}, \cite{Damour85}):
        \begin{equation*}
            \Delta\omega = \frac{6\pi R_g}{a(1-e^2)}
        \end{equation*}
        Dividing this value by \emph{synodic period $T$} (the time that it takes for the probe body to go from pericenter to pericenter) gives us the value of the rate of the pericenter shift:
        \begin{equation}
        \label{GR-pericenter-shift}
            \dot{\omega} = \frac{6\pi R_g}{a(1-e^2)T}
        \end{equation}
        
\subsection{In Field Gravitation Theory}
    In the frame of the \emph{field approach to gravitation}, formulated by Feynman in his famous "\textbf{Lectures on Gravitation}" \cite{FL}, the gravitation  is described as a \emph{material field with positive energy density} in a flat Minkowski space (like all other fundamental physical interactions).\\
    The gravitational symmetric tensor field has two irreducible parts: spin-2 attractive part and spin-0 repulsive part (V. V. Sokolov, Yu. V. Baryshev, \cite{sb1980}, 1980). One of the main characteristics of the field is that it has energy-momentum tensor, and  its 00-component (energy density of gravitational field) from \cite{sb1980} for static spherically symmetric field is given by:
    \begin{equation}
    \label{FG-grav-energy}
        \boxed{t^{00}_g = \frac{1}{8\pi G}(\bm{\nabla}\varphi_N)^2 \geqslant 0}
    \end{equation}
    \subsubsection{Problem of the energy-momentum pseudotensor in GR}
        According to Landau and Lifshitz ``The Classical Theory of Fields" (\cite{LL2}, §96) the energy-momentum of the gravitational field in curved space cannot be defined in covariant form. It is so called the  energy-momentum pseudotensor of gravitational field, and it depends on a choice of coordinates, so: \emph{"it is meaningless to talk of whether or not there is gravitational energy at a given place"}. Thus, the energy of gravitational field is \textbf{non-localizable} in the frame of geometrical approach for description of gravitation.\\
    While in the Feynman's field approach to gravitation, the energy density of the gravitation field is a positive localizable physical quantity, e.g.
    Eq.(\ref{FG-grav-energy}).
    \subsubsection{The equations of motion (FGT)}
        In the frame of the Field Gravitation Theory, according to Baryshev (\cite{b1986}, 1986) (see also G. Kalman, \cite{Kalman61}, 1961), the Lagrange function and corresponding Post-Newtonian equations of motion of a test particle in the gravitational field of central massive body is given by:	
        \begin{equation*}
            \frac{L}{m} = \frac{\dot{\mathbf{r}}^2}{2} \bigg( 1 + \frac{\dot{\mathbf{r}}^2}{4c^2} - 3\frac{\varphi_N}{c^2} \bigg) - \varphi_N \bigg( 1 + \frac{\varphi_N}{2c^2} \bigg)
        \end{equation*}
        \begin{equation}
        \label{FGT-PN-eq-mot}
            \boxed{\ddot{\mathbf{r}}=-\bm{\nabla}\varphi_N \bigg( 1 + 4\frac{\varphi_N}{c^2} + \frac{\dot{\mathbf{r}}^2}{c^2} \bigg) + 4 \bigg( \bm{\nabla}\varphi_N \cdot\frac{\dot{\mathbf{r}}}{c} \bigg)\frac{\dot{\mathbf{r}}}{c}}
        \end{equation}\\
        We see that the Eq.(\ref{FGT-PN-eq-mot}) coincide with the equation Eq.(\ref{GR-harmon}) that we have obtained in the frame of harmonic coordinates of GR.

    \subsubsection{Pericenter shift (FGT)}
        From coincidence of equations of test particles motion
        Eq.(\ref{FGT-PN-eq-mot}) in FGT and Eq.(\ref{GR-harmon}) in GR it follows that the expression for the pericenter shift in FGT and GR is exactly the same.\\
        The difference from GR is that the FGT formula has two different terms:
        \begin{equation*}
            \dot{\omega} = \frac{6\pi R_g}{a(1-e^2)T} = \frac{7\pi R_g}{a(1-e^2)T} - \frac{\pi R_g}{a(1-e^2)T}
        \end{equation*}
        The first term with a factor of $7\pi$ corresponds to the linear approximation, when in the field equations one does not take into account the non-linear contribution (energy of gravitational field itself).\\
        The second term with a factor of $-\pi$ occurs after taking into account the non-linearity due to the positive energy density of the gravitational field given by Eq.(\ref{FG-grav-energy}). This term corresponds to the measuring the field energy of the gravitation via pericenter shift observations.
\subsection{Pericenter shift values of S-stars}
    The pericenter shift values and parameter of $\frac{\varphi_N}{c^2}$ in pericenter predicted for S-stars are:
    \begin{table}[H]
        \centering
        \begin{tabular}{lccc}
            \hline
            Star            & S2 & S38 & S55 \\ \hline
            $\Delta\omega$  & $12'$ & $7.1'$ & $6.7'$ \\
            $\dot{\omega}$  & $45''\textrm{/yr}$ & $22''\textrm{/yr}$ & $31''\textrm{/yr}$ \\
            $\dot{\omega}\cdot100\;\textrm{yrs}$ & $1^\circ15'$ & $37'$ & $52'$ \\
            $\big(\frac{\varphi_N}{c^2}\big)_{\textrm{per}}$ & $-3.48\cdot10^{-4}$ & $-1.99\cdot10^{-4}$ & $-1.69\cdot10^{-4}$ \\ \hline
        \end{tabular}
        \caption{S-stars pericenter shift}\label{tab:Sstars-pericenter}
    \end{table}
    \noindent The observable effect is similar to one with \textbf{Mercury anomalous pericenter shift}, that Einstein explained. Now we can observe it with S-star cluster in the center of our Galaxy.\\
    For Mercury, the pericentral $\frac{\varphi_N}{c^2}$ is $-3.21\cdot10^{-8}$, and for binary pulsar PSR 1913+16 it is $-2.7\cdot10^{-6}$. We can see that the field at the pericenter of the S-stars orbits is stronger by 2 orders of magnitude. Although, it is still considered as a weak field.
    
\section{Simulations}
    We have applied 4-order Runge-Kutta integrator to solve the equations of motion Eq.(\ref{GR-harmon}) (Eq.(\ref{FGT-PN-eq-mot})) to simulate the problem of the Post-Newtonian motion of a test particle in a gravitational field of central massive body.\\
    The conservation laws were used to obtain pericenter velocities by given semi-major axis and eccentricity in Newtonian and Post-Newtonian cases (see the details in my future work).\\
    \subsection{S2 star orbit PN effects}
    \subsubsection{Orbit simulation}
        We have substituted S2 star parameters into the integrator to simulate its orbit. We obtained the observable trajectory:\\
        \begin{minipage}{0.6\linewidth}
            \centering
            \includegraphics[scale=0.62, trim={1.0cm 0.5cm 1.2cm 0.9cm}, clip]{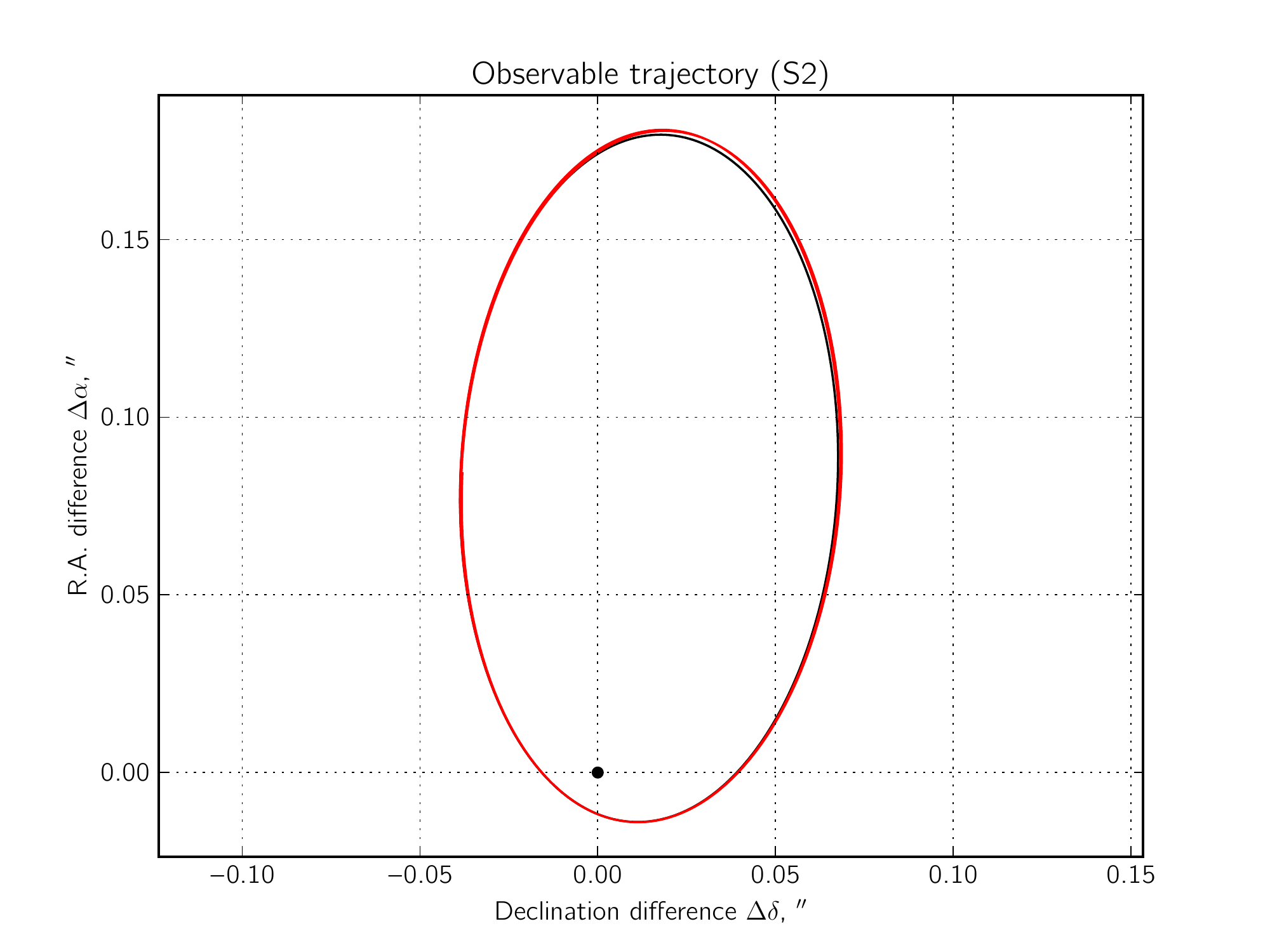}
        \end{minipage}
        \begin{minipage}{0.06\linewidth}
            $\;$
        \end{minipage}
        \begin{minipage}{0.33\linewidth}
        \fussy
            \noindent The upper part of the trajectory, zoomed in:
            \begin{center}
                \fbox{\includegraphics[scale=1.35]{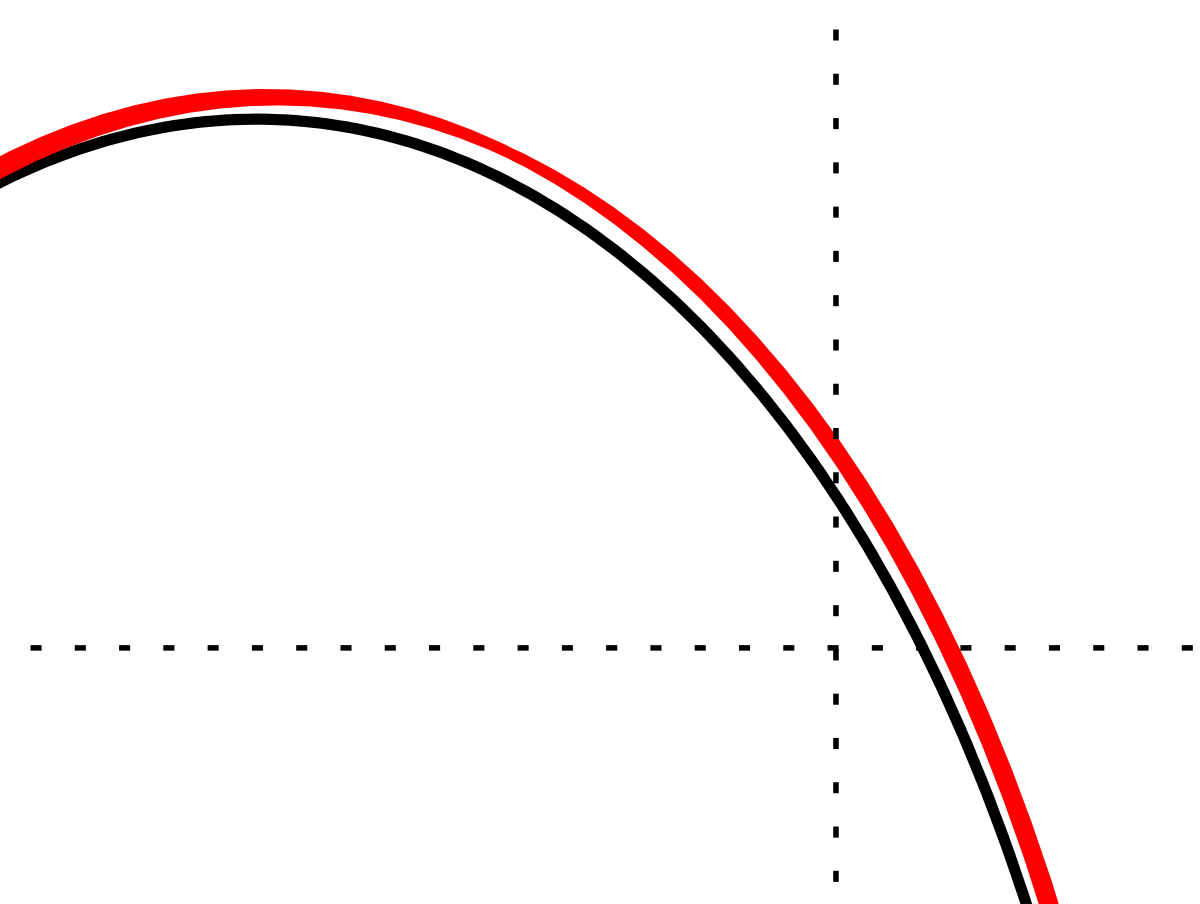}}
            \end{center}
            We can see that \textcolor{red}{\textbf{Post-Newtonian}} orbit is slightly thicker than \textbf{Newtonian}. That is caused by the pericenter shift effect: there are several slightly different passages, that stacked on top of each other.
        \end{minipage}

    \subsubsection{Redshift: Relativistic Doppler effect, Gravitatonal redshift, and total}
    
    \begin{minipage}{0.5\linewidth}
        The difference between the Newtonian and Post-Newtonian equations of motions leads into the small difference in the pericenter velocities, which leads into the small difference in orbital periods. We can see that after approximately 30 years of observations the difference is becoming larger due to time offset.
        \vspace{0.3cm}\\
        The same effect can be seen on gravitational redshift plot.
        \vspace{0.3cm}\\
        We have also obtained total redshift plot, which takes into accound both Doppler shift and gravitational redshift.
    \end{minipage}
    \begin{minipage}{0.5\linewidth}
        \centering
        \includegraphics[scale=0.5, trim={0.5cm 0.4cm 1.2cm 0.8cm}, clip]{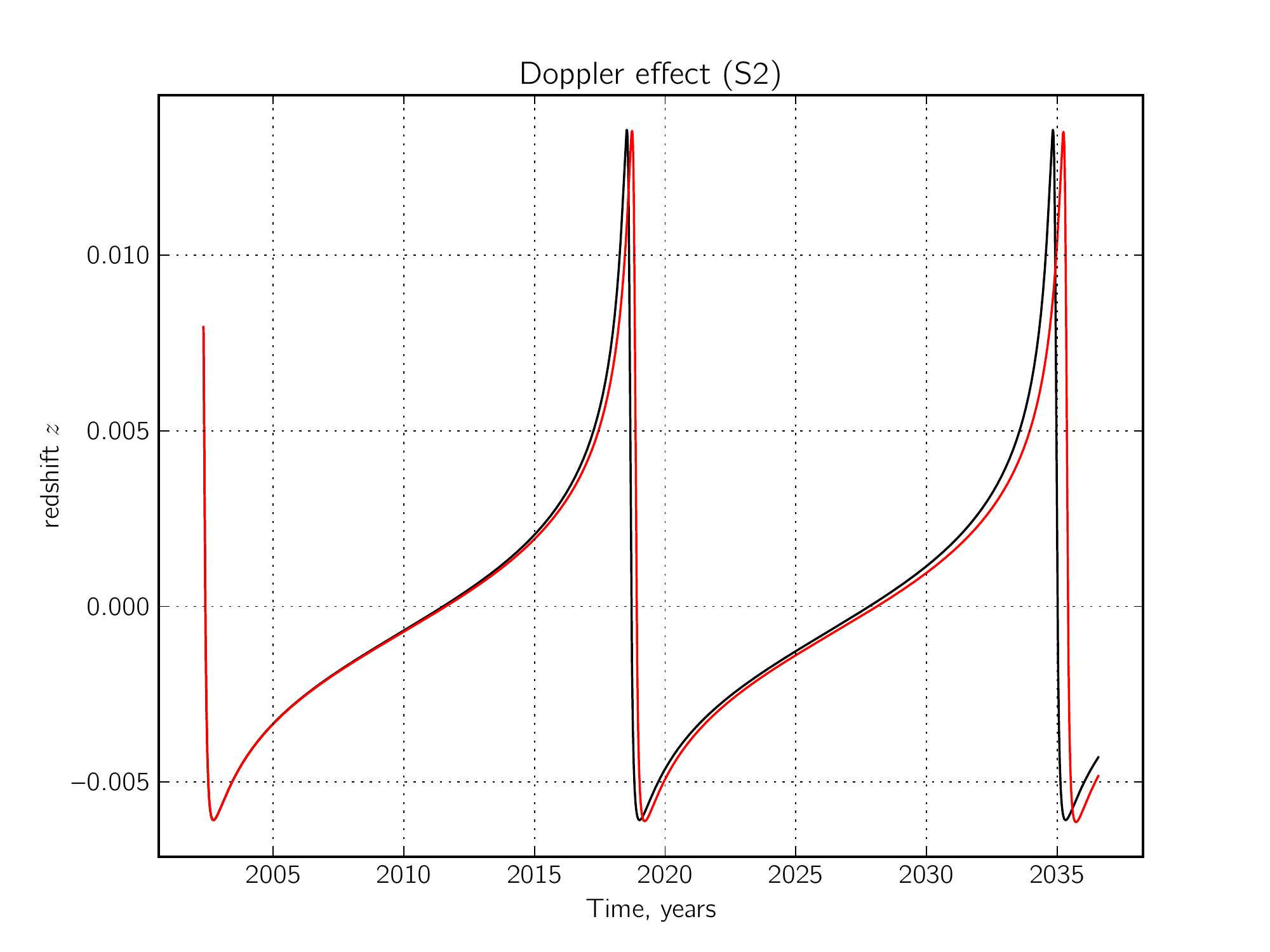}
    \end{minipage}\\
    \begin{minipage}{0.5\linewidth}
        \centering
        \includegraphics[scale=0.5, trim={0.5cm 0.4cm 1.2cm 0.8cm}, clip]{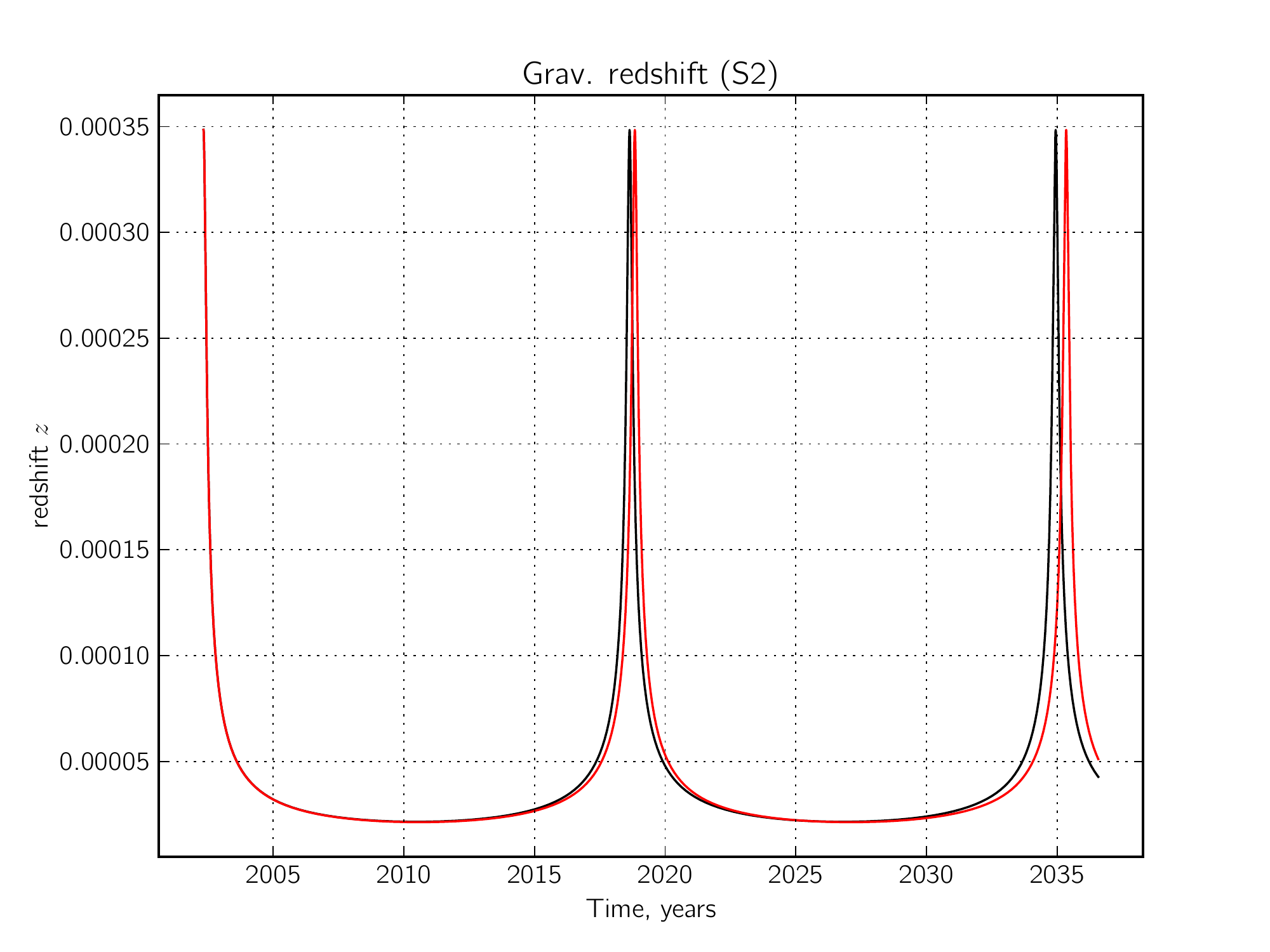}
    \end{minipage}
    \begin{minipage}{0.5\linewidth}
        \centering
        \includegraphics[scale=0.5, trim={0.5cm 0.4cm 1.2cm 0.8cm}, clip]{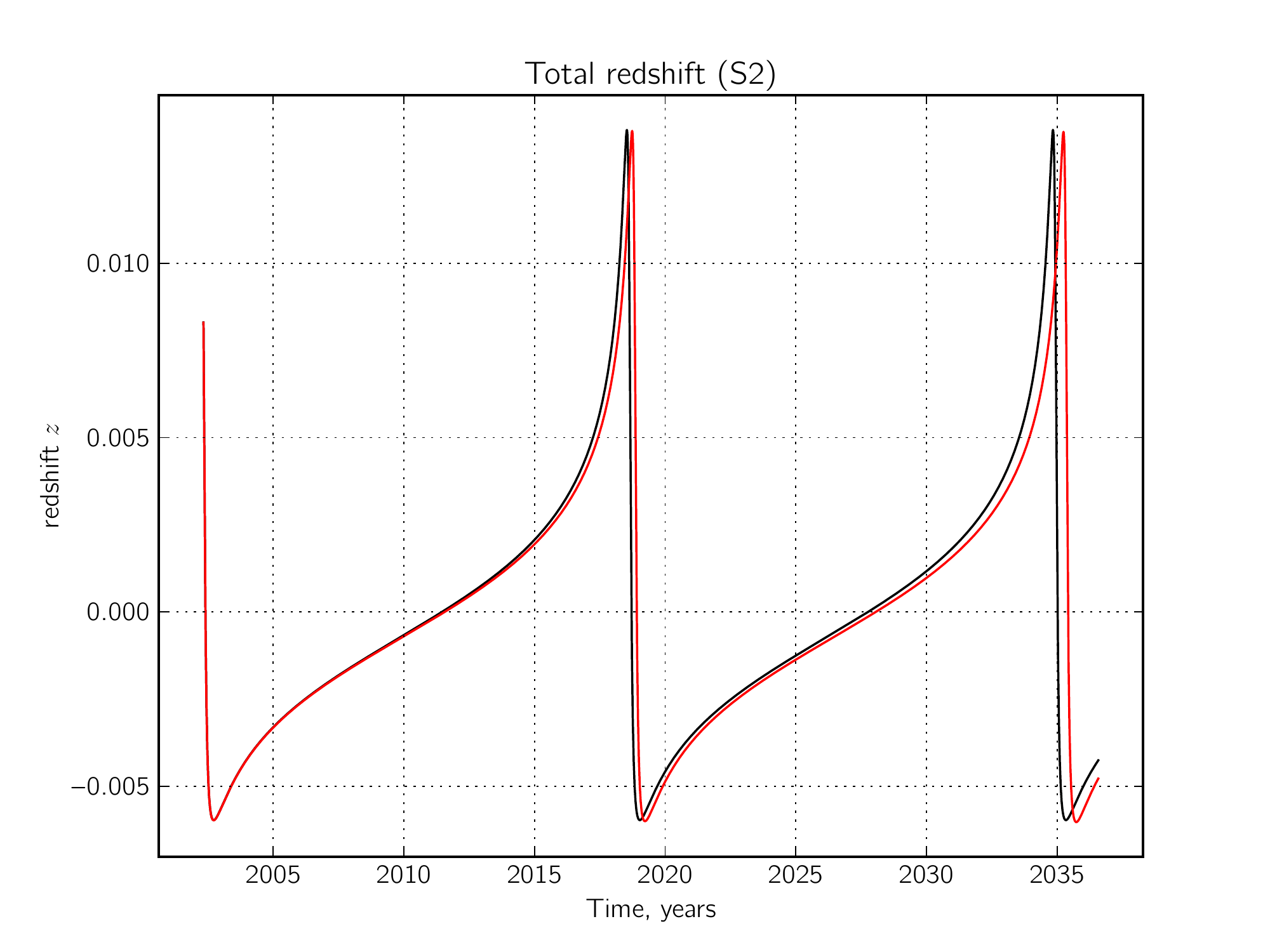}
    \end{minipage}
    
    \subsection{Energy density of the gravitational field}
    \begin{minipage}{0.4\linewidth}
        According to the Eq.(\ref{FG-grav-energy}) we also can measure the energy density of the gravitational field along the orbit of a star:
        \begin{equation*}
            \varepsilon_g = \frac{(\bm{\nabla}\varphi_N)^2}{8\pi G} = \frac{GM^2}{8\pi r^4}
        \end{equation*}
    \end{minipage}
    \begin{minipage}{0.6\linewidth}
        \centering
        \includegraphics[scale=0.6, trim={0.5cm 0.4cm 1.2cm 0.8cm}, clip]{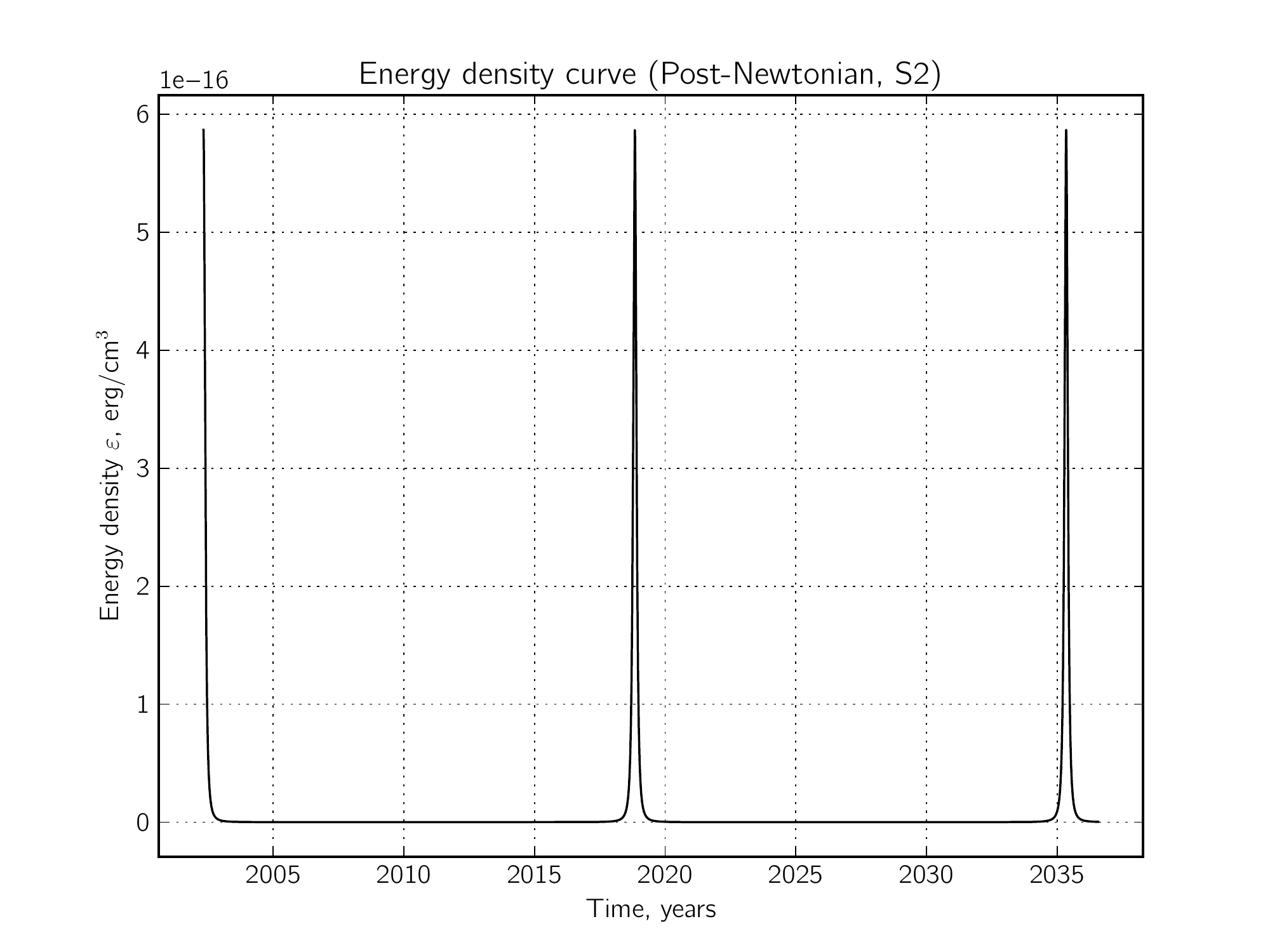}
    \end{minipage}

\section{Conclusion}
    The predicted for observations Post-Newtonian effects for the nearest to Sgr A* stars coincide in GR and FGT (Field Gravitation Theory).
    \vspace{0.3cm}\\
    In the frame of the FGT the effect of pericenter shift
    contains separate term, which corresponds to the direct contribution of the \textbf{positive localizable energy density} of the graviational field.  
    \vspace{0.3cm}\\
    The value of positive energy density of gravitational field in the frame of field approach to gravitation is measurable. This conclusion is also consistent with the detection of positive energy of the gravitational waves by LIGO-Virgo antennas.

\end{document}